\documentclass[lettersize,journal]{IEEEtran}
\usepackage{amsmath,amsfonts}
\usepackage{algorithmic}
\usepackage{algorithm}
\usepackage{array}
\usepackage[caption=false,font=normalsize,labelfont=sf,textfont=sf]{subfig}
\usepackage{textcomp}
\usepackage{stfloats}
\usepackage{url}
\usepackage{verbatim}
\usepackage{graphicx}
\usepackage{cite}
\usepackage[hidelinks]{hyperref}
\usepackage{lipsum} 
\usepackage{textcomp}
\usepackage{listings}
\usepackage[center]{caption}
\usepackage{gensymb}

\renewcommand{\baselinestretch}{0.96}

\begin{document}

\title{HULK-V: a Heterogeneous Ultra-low-power Linux capable RISC-V SoC}

\author{\IEEEauthorblockN{Luca Valente\IEEEauthorrefmark{1},  Yvan Tortorella \IEEEauthorrefmark{1}, Mattia Sinigaglia \IEEEauthorrefmark{1}, Giuseppe Tagliavini\IEEEauthorrefmark{1}, Alessandro Capotondi\IEEEauthorrefmark{2}, Luca Benini\IEEEauthorrefmark{1}\IEEEauthorrefmark{3}, Davide Rossi\IEEEauthorrefmark{1}} \\
\IEEEauthorblockA{\textit{DEI, University of Bologna, Italy}\IEEEauthorrefmark{1}  \textit{University of Modena and Reggio Emilia}\IEEEauthorrefmark{2} \textit{IIS lab, ETH Zurich, Switzerland}\IEEEauthorrefmark{3} \\ $\{$name.surname$\}$@unibo.it} \vspace{-7mm}} 
\maketitle

\begin{abstract}
IoT applications span a wide range in performance and memory footprint, under tight cost and power constraints. High-end applications rely on power-hungry Systems-on-Chip (SoCs) featuring powerful processors, large LPDDR/DDR3/4/5 memories, and supporting full-fledged Operating Systems (OS). On the contrary, low-end applications typically rely on Ultra-Low-Power $\mu$controllers with a "close to metal" software environment and simple micro-kernel-based runtimes.
Emerging applications and trends of IoT require the "best of both worlds": cheap and low-power SoC systems with a well-known and agile software environment based on full-fledged OS (e.g., Linux), coupled with extreme energy efficiency and parallel digital signal processing capabilities.
We present HULK-V: an open-source Heterogeneous Linux-capable RISC-V-based SoC coupling a 64-bit RISC-V processor with an 8-core Programmable Multi-Core Accelerator (PMCA), delivering up to 13.8 GOps, up to 157 GOps/W and accelerating the execution of complex DSP and ML tasks by up to 112$\times$ over the host processor. HULK-V leverages a lightweight, fully digital memory hierarchy based on HyperRAM IoT DRAM that exposes up to 512 MB of DRAM memory to the host CPU. Featuring HyperRAMs, HULK-V doubles the energy efficiency without significant performance loss compared to featuring power-hungry LPDDR memories, requiring expensive and large mixed-signal PHYs. HULK-V, implemented in Global Foundries 22nm FDX technology, is a fully digital ultra-low-cost SoC running a 64-bit Linux software stack with OpenMP host-to-PMCA offload within a power envelope of just 250 mW.
\end{abstract}


\vspace{-5mm}
\section{Introduction}\label{introsect}
As we entered a new Internet of Things era, the number of IoT devices and the spectrum of IoT applications are continuously increasing: from home applications, robotics, industrial gateways, drones, and building automation to smart cities, digital signage, medical equipment, and more~\cite{iot_review}. 

Depending on the application's requirements, commercial IoT devices can either be high-end computing platforms, called Single Board Computers (SBCs), or low-end $\mu$controllers (MCUs) with ultra-low-power consumption. To keep the power envelope within 200 mW, MCUs usually feature simple RISC host processors (e.g., ARM Cortex-M) to which they expose just a few hundred KBytes of on-chip SRAM memory, and leverage heterogeneity to achieve high data processing capabilities\cite{PULP}. Due to the limited amount of memory, MCUs can only support lightweight real-time Operating Systems (OS) or custom bare-metal runtimes. On the other hand, to support full-fledged OSs (e.g., Linux) and provide access to high-level software libraries, SBCs feature application-class cores, like multi-core ARM Cortex-A processors\cite{pi0}, GPUs, and GBytes of high-performance off-chip LPDDR/DDR/3/4/5 memories, with dedicated large (few $mm^2$ in 22 nm node~\cite{xeon}), mixed-signal, proprietary and expensive ($>300$ thousands dollars\cite{stinky}) on-chip PHY controllers, ending up with a power envelope of few Watts. 

Even though an increasing number of applications require low power consumption, low silicon cost and highly energy-efficient data processing capabilities coupled with standard and mature programming interfaces, there is still no hardware platform offering all these characteristics simultaneously. This work presents HULK-V: a Heterogeneous Ultra-Low-power Linux-ready RISC-V SoC, implemented in Global Foundries 22nm FDX technology. HULK-V's heart is the 64-bit RISC-V Linux-capable CVA6~\cite{zaruba2019cost} core, accelerated by an ultra-low-power energy-efficient programmable cluster composed of 8 32-bit RISC-V cores enhanced with integer and floating-point DSP extensions. The cluster can deliver up to 13.8 GOps and 157 GOps/W over a broad spectrum of IoT applications. HULK-V supports an OpenMP-based user-space library that provides an intuitive programming interface to offload parallel code from the host to the accelerator.

To fit the tight power and cost requirements of many IoT applications while exposing hundreds of MB to the host processor, HULK-V replaces power-hungry LPDDR memories and large mixed-signal PHY controllers with HyperRAMs~\cite{hyperram_low_pincount} and an ultra-low-power cheap 0.27$mm^2$ fully-digital memory controller. HyperRAMs belong to the family of IoT memories, like RPCDRAMs~\cite{rpc}, providing relatively high-bandwidth, low-pin count, ease of integration, low power consumption, and enough memory capacity to run a full-fledged OS like Linux. To mitigate the performance impact of the lower bandwidth of such memories compared to LPDDRs, HULK-V integrates a Last-Level Cache (LLC) tightly coupled to the memory controller.

Featuring HyperRAMs, HULK-V increases the energy efficiency by up to 2$\times$ on IoT ML applications, with negligible performance penalty on CPU-centric benchmarks compared to featuring LPDDR4 memories, usually needed to run full-fledged OSs. To the author's knowledge, HULK-V is the only Linux-capable, heterogeneous, and programmable platform, delivering GOps range performance at extremely high energy efficiency within a power envelope of only 250 mW and die area smaller than 9$mm^2$. The hardware and software described in this work are open-source, intending to support and boost an innovation ecosystem focusing on ULP computing for the IoT landscape.

\vspace{-3mm}
\section{Related work}\label{related}

IoT devices can be classified into three categories: low-end, mid-end, and high-end devices\cite{iot_review}. Low-end devices, like \cite{arduinonano}, typically run at very low frequency ($<$30 MHz), have minimal on-board resources ($<$100 kB of on-chip SRAM) and connectivity, and consume a few mW of power, with minimal software support. In this work, we rather focus on mid-end and high-end devices with moderate performance and support for a full-fledged OS. Table~\ref{tab:related} shows the systems closest to our approach, namely \emph{mid-end} and \emph{high-end} devices.

\vspace{-3mm}
\subsection{Mid-end devices}
\emph{Mid-end} devices, like~\cite{PULP}\cite{sapphire}\cite{crossover}, usually have a simple host core and a few hundreds of kB of on-chip scratchpad SRAM and tens of MB of off-chip DRAM. Due to the limited memory resources, these devices provide only support for lightweight RTOSs and custom runtimes. The off-chip DRAM is accessible only through IOs (SPI, QSPI, ...), and all transactions are explicitly managed through drivers offering in/out copy APIs. State-of-the-art mid-end devices have high energy-efficient DSP capability, thanks to ISA optimizations or dedicated hardware accelerators.

Vega~\cite{PULP} is a mid-end IoT platform consuming less than 100 mW. It comprises a host RISC-V core and an accelerator featuring 9 RISC-V cores optimized for DSP reduced precision computation, supporting integer and FP arithmetic. Vega also features a neural networks hardware accelerator, delivering up to 32 GMAC/s on int8 data, and it integrates an HyperBUS controller.  Unlike HULK-V, Vega does not directly expose the off-chip DRAMs to the host processor and only supports RTOS and bare-metal runtimes: FreeRTOS and PULP-OS.

Sapphire~\cite{sapphire} is a soft-SoC available for FPGA deployment. The off-chip memory is memory-mapped on the AXI interconnect, and customers can choose between DDR or HyperRAM memories on board. Differently from HULK-V, Sapphire is not available as ASIC, and its processor does not support application-class features, like virtual memory or multiple privilege levels, necessary to run a Linux OS. 

i.MX RT1180~\cite{crossover} is an embedded SoC from NXP featuring a powerful ARM Cortex-M7, reaching up to 800 MHz. The SoC provides the CPU with a big 1.5 MB on-chip memory, but external memories are not directly accessible by the core, limiting the memory mapped footprint available for OSs. Moreover, the ARM Cortex-M7 does not provide Linux support, and it delivers much less operational throughput than HULK-V.

\vspace{-4mm}
\subsection{High-end devices}

\emph{High-end} devices, like~\cite{unmatched}\cite{Jetson}\cite{pi0}, feature more powerful CPUs, running at more than 1 GHz, and hundreds of MB of memory provided by LPDDR3/4/5 devices and their PHY controllers, which consume more than one Watt~\cite{powerhighend}. High-end SoCs might also have integrated GPUs\cite{Jetson}. These devices are known as \textit{Single Board Computers} (SBC) thanks to their capability of running complete Linux OS distribution.

Pi0~\cite{pi0} from Raspberry is probably the most popular SBC on the market, and it is the archetype of a simple SBC with a significantly reduced form factor. It is powerful but power-hungry, with its Quad-core ARM Cortex-A53 running at 1 GHz and 512 MB of LPDDR2 as main memory. The same reasoning applies to Nvidia's Orin~\cite{Jetson} and SiFive's Unmatched~\cite{unmatched}. 

HULK-V joins the benefits of SBC and MCU systems. The main CPU memory comprises HyperRAMs, not oversized DDRs or small on-chip SRAMs. Such a setup allows the processor to run Linux while maintaining a limited power consumption and a simplified form factor. Furthermore, thanks to the accelerator cluster, HULK-V delivers state-of-the-art computing performance compared to mid-end embedded systems, associated with a productive and mature programming environment based on a complete SPM Linux Distribution coupled with an intuitive and effective heterogeneous programming interface based on OpenMP 5.

\begin{table}[]
    \centering
    \caption{Comparison with State-of-Art}
    \begin{tabular}{|c|c|c|c|c|c|} \hline
                                      & OS    & Memory             & ASIC/     & Host      & Accel-   \\ 
                                      &       &                    & FPGA      & CPU       & erators  \\ \hline
         Vega                         & RTOS  & 512KB SRAM         & ASIC      & Ri5cy     & PMCA     \\
         \cite{PULP}                  &       & 512MB Hyper        &           & 200MHz    &          \\ \hline
         Sapphire                     & RTOS  & 4MB-3GB            & FPGA      & VexRiscv  & No       \\ 
         \cite{sapphire}              &       & DDR/Hyper          &           & 400MHz    &          \\ \hline
         i.MX RT                      & RTOS  & 1.5MB              & ASIC      & CortexM7  & MIPI     \\  
         \cite{crossover}             &       & SRAM               &           & 800MHz    &          \\ \hline 
         HeroV2                       & Linux & 1GB                & FPGA      & Quad-Core & PMCA     \\  
         \cite{herov2}                &       & DDR4               &           & CortexA53 &          \\  
                                      &       &                    &           &  1GHz     &          \\ \hline 
         Raspberry                    & Linux & 512MB              & ASIC      & Quad-Core & No       \\ 
         Pi0 \cite{pi0}               &       & LPDDR2             &           & CortexA53 &          \\ 
                                      &       &                    &           & 1GHz      &          \\ \hline
         Unmatched                    & Linux & 16GB               & ASIC      & U74       & No       \\ 
         \cite{unmatched}             &       & DDR4               &           & 1GHz      &          \\ \hline
         This                         & Linux/ & 512KB SRAM         & ASIC/     & CVA6      & PMCA     \\ 
         work                         & RTOS  & 512MB Hyper        & FPGA      & 900MHz    &          \\ \hline
    \end{tabular}
    \label{tab:related}
    \vspace{-5mm}
\end{table}

\vspace{-4.5mm}
\subsection{Research platforms}
When compared to academia, several projects could resemble our approach, especially regarding heterogeneity. Three research projects that aim to build heterogeneous SoC, with a particular focus on cache coherency, are ESP~\cite{ESP}, BYOC~\cite{BYOC}, and AGILER~\cite{kamaleldin2022agiler}. Platforms like ESP, BYOC, and AGILER are research platforms used for the architectural exploration of many-core high-performance systems. They rely on DDR memories, many application-class cores, and accelerators to meet their performance target. On the contrary, HULK-V features only one application-class core, and it is sized for embedded applications, hitting a much smaller power envelope. 

The closest one is HERO~\cite{herov2}. HERO is an FPGA-based research platform developed to enable accurate and fast exploration of heterogeneous computers consisting of an 8 RISC-V cores cluster and an ARM Cortex-A53 host processor. While the asymmetry of the architecture is similar to ours, the target is entirely different. In HERO, the host system is the Zynq-SoC running at 1.2 GHz, and the PULP cluster is only emulated at 50 MHz, not providing any actual speed up against the host. On the contrary, HULK-V is a deeply optimized ASIC SoC, it is not an FPGA platform for architectural exploration and the PMCA actually speeds up the execution of parallel workloads against the host. Furthermore, differently from HERO, all IPs integrated in HULK-V are open-source, except for the technology dependent ones.

\vspace{-3mm}
\section{Architecture}\label{ARCH}
\begin{figure}[!t]
    \centering
    \includegraphics[width=.5\textwidth]{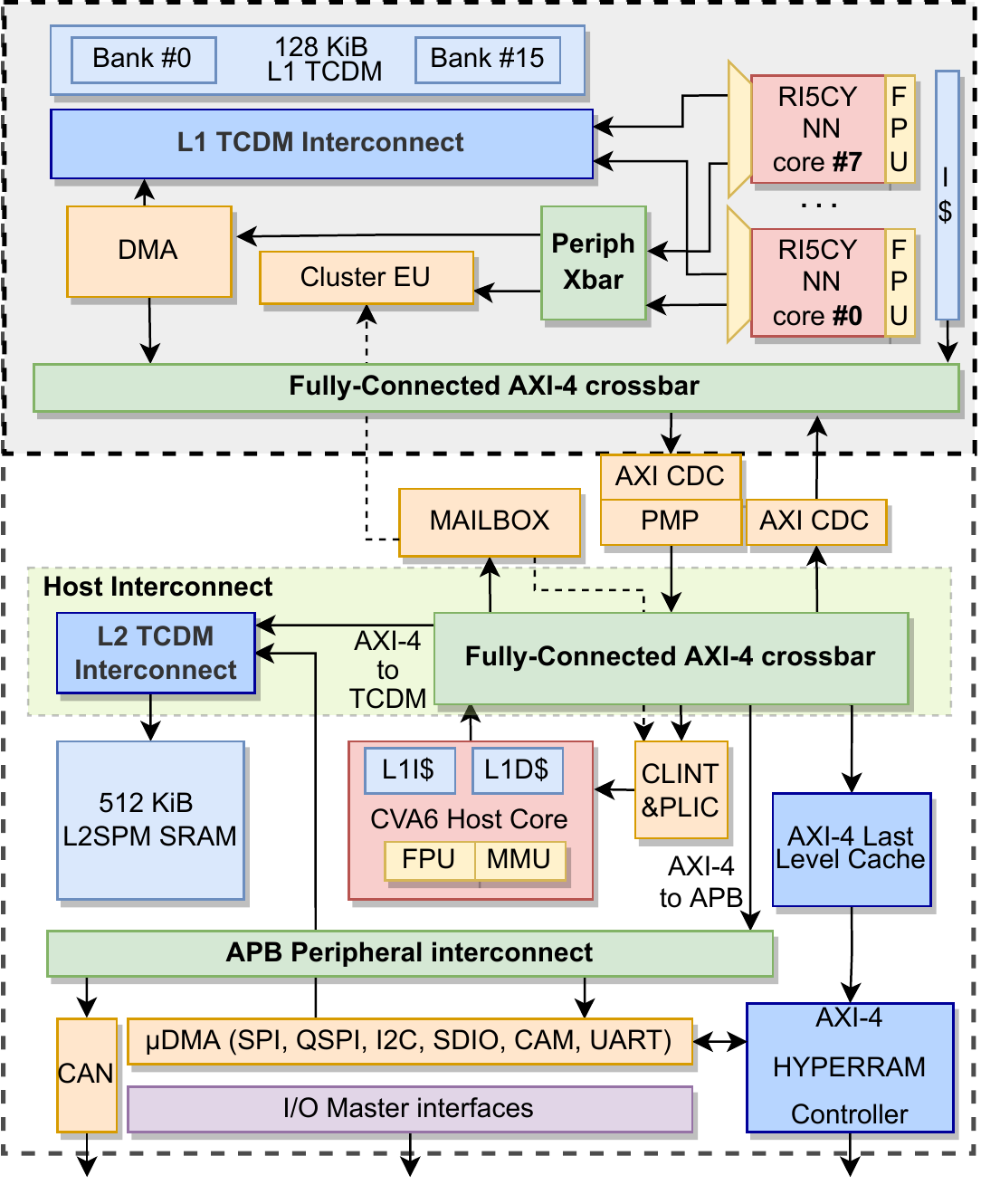}
    \caption{Overall HULK-V SoC Architecture.}
    \label{fig:uarch}
    \vspace{-4mm}
\end{figure}

This section describes the heterogeneous system architecture focusing on the two key elements of the system: i) the low-power and cheap memory system and ii) the parallel programmable accelerator. Figure~\ref{fig:uarch} contains the block diagram of HULK-V. The SoC is divided into four frequency domains, adjusted by four Frequency Locked Loop (FLL): the host core, the host domain, the peripheral domain, and the accelerator cluster.

The heart of the host subsystem is the CVA6~\cite{zaruba2019cost} core. CVA6 is a 6-stages, single-issue, in-order, 64-bit RISC-V core, supporting the RV64GC ISA variant, virtual memory, three execution privilege levels, physical memory protection (PMP), and the Linux OS. CVA6 has 16 kB of L1 I-cache and 32 kB of write-through L1 D-cache to enable simple coherency with other masters to the interconnect.

The main host interconnect is a high-bandwidth, low-latency 64-bit AXI4 crossbar. The host domain provides a 512 kB scratchpad memory (L2SPM) and a complete set of peripherals (I2C, (Q)SPI, CPI, SDIO, UART, CAN, PWM, I2S) serving the requirements of IoT applications in several fields, such as automotive, audio, and robotics. Data to/from off-chip peripherals are autonomously written/read from/to the L2SPM through a dedicated $\mu$DMA. The host domain is also composed of a standard Platform Level Interrupt Controller (PLIC), a Core Local Interrupt (CLINT), the HyperRAM controller, and an LLC.

\vspace{-4mm}
\subsection{Last Level Cache}
\begin{figure}[t]
    \centering
    \includegraphics[width=0.5\textwidth]{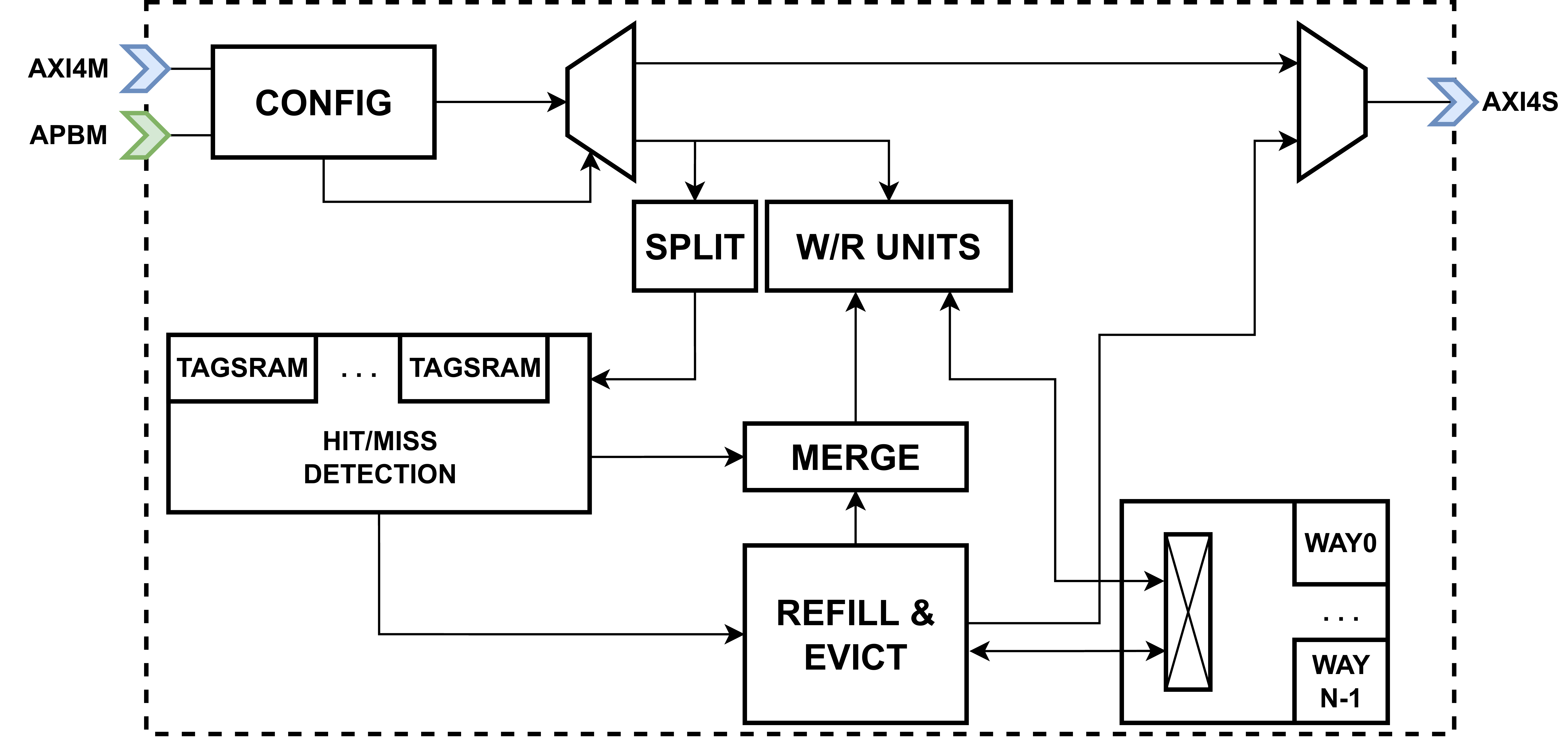}
    \caption{Last Level Cache Architecture.}
    \label{fig:LLC}
    \vspace{-2mm}
\end{figure}

CVA6 needs a Last Level Cache (LLC) to cope with the intrinsic high latency of the HyperRAMs and to deliver maximum performance. The LLC's architecture is shown in Figure~\ref{fig:LLC}. Incoming AXI transactions are first filtered. The requests inside the cacheable region are passed to the cache, while the others are directly propagated out to the external memory.

Cacheable AXI transactions are then split onto the descriptors used in the hit/miss tag-lookup mechanism. Tags are stored in SRAM and are accessible in one clock cycle. After the tag-lookup, the descriptor goes directly to the read/write units on a hit.
On a miss, a cache line must be evicted and refilled. An eviction generates an AXI write transaction on the output port through the read unit. Then, the refill triggers an out AXI read transaction on the output port through the write unit. 

The HULK-V's LLC design is highly parameterizable. "Blocks" are as wide as the AXI data width ($AXI_{dw}$). Then, one can choose the number of blocks in a cache line ($N_{blocks}$), the number of lines in a set ($N_{lines}$) and the number of ways ($N_{ways}$). The resulting LLC size will be equal to: $LLC_{size} = N_{ways} \cdot N_{lines} \cdot N_{blocks} \cdot AXI_{dw}$. In HULK-V, we set the number of blocks to 8, the number of lines to 256, and the number of ways to 8, which means 128 kB of LLC.

\vspace{-3mm}
\subsection{Low-cost, low-power HyperRAM controller}
\begin{figure}
    \centering
    \includegraphics[width=.5\textwidth]{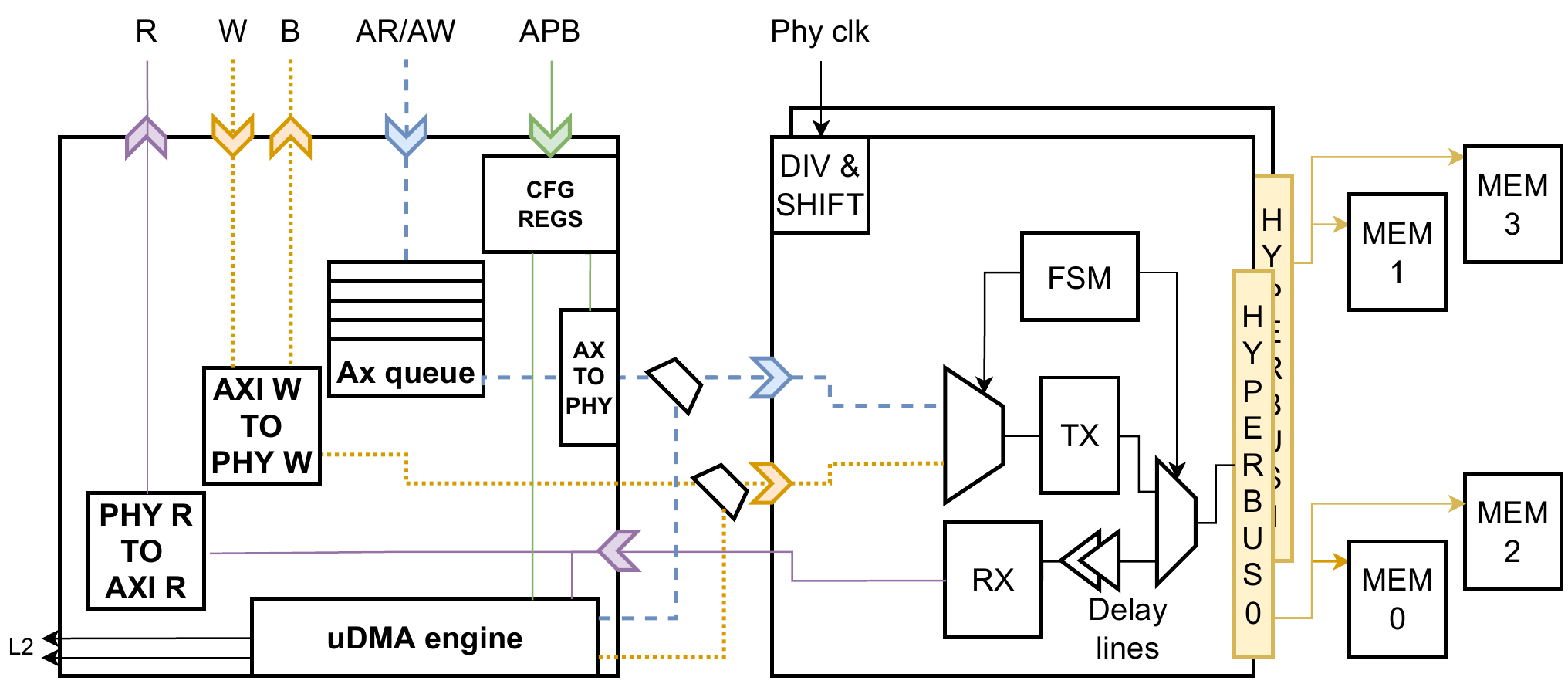}
    \caption{HyperRAM Memory Controller Architecture.}
    \label{fig:hyper}
    \vspace{-2mm}
\end{figure}

HULK-V's HyperRAM controller is depicted in Figure~\ref{fig:hyper}. It provides an AXI4 slave port and a configuration APB port; it connects the SoC with the off-chips HyperRAMs. The HyperBUS protocol is a fully digital protocol counting $11+n$ pins: $3$ control pins, $n$ Chip Select (CS), and $8$ Double-Data-Rate pins. Latest HyperRAMs can reach up to 200 MHz and provide up to 3.2 Gbps and up to 64 MB of capacity\cite{hyperram_low_pincount}.

The HyperRAM controller comprises two modules: the PHY controller and the front-end, in separate frequency domains. The front-end contains an AXI4-to-PHY converter and a dedicated $\mu$DMA engine programmable through APB for software-programmed DMA transfers. The AXI and $\mu$DMA transactions are multiplexed towards the PHY, which translates the incoming data packets into HyperRAM transactions and vice versa.

The AXI front-end enqueues the AXI transactions one at a time, and it translates the oldest pending AR/AW transaction into a data packet for the PHY. Then, if it is a write, the W channel transactions get converted into multiple PHY data packets. For reads, the mechanism is the same, with the PHY back-end sending data packets to the converter that then populates the R channel.

The $\mu$DMA engine directly connects the L2SPM and the HyperRAM and can generate both 1D and 2D burst transactions. Such features are precious for efficiently executing ML algorithms on the cluster. Doing so requires careful tiling of the input weights: filling the L2SPM with as many weights as possible and then bringing a smaller portion of them into the L1SPM, accessible by the cores in one clock cycle.

The proposed module is highly parameterizable. One can choose how many memories to connect on the same HyperBUS to set the overall available memory. Multiple memories on the same bus are placed contiguously in the address memory map and selected through their dedicated CS. At runtime, one can communicate to the controller the size of the HyperRAMs, and the controller will demultiplex the transactions accordingly.

Also, one can choose how many HyperBUS interfaces (1 or 2) to expose. Both buses will have the same number of CSs. When exposing 2 HyperBUSes, the pair of memories on the same chip select will be mapped as interleaved: each memory will be seen as a memory block of 16-bit width. Doing so will double the maximum achievable bandwidth, up to 6.4 Gbps, doubling the pin count.

\vspace{-2mm}
\subsection{Programmable Multi-Core Accelerator}

The Programmable Multi-Core Accelerator (PMCA) is built around 8 CV32E4-based cores~\cite{ri5cy} sharing 16$\times$8 kB SRAM banks, composing a 128 kB L1SPM, and it is invoked by the host to run computation-intensive kernels. The cores feature a custom RV32 extension, providing many DSP and ML features, like hardware loops, MAC\&Load operation, SIMD operations, and post-increment LD/ST. The cluster features 8 FPUs supporting FP32 and FP16 with SIMD support. SIMD operations, not available in the CVA6 host core, reduce the operands' width to double or quadruple the number of operations per cycle. On integer numbers, the precision can be reduced down to 8-bit and down to 16-bit for FP. The cluster also features a two-level I-cache: 512 B for each core and 4 kB shared.

The cluster architecture is optimized for embedded systems and ML algorithms. To avoid the hardware overhead of data caches, the cluster exploits scratchpad memories, DMA accesses, double-buffering, and custom ISA extension to maximize the utilization of memory and computing resources through explicit memory management of the memory~\cite{dory}.

The PMCA communicates with the host's AXI4 interconnect through two 64-bit AXI4 ports, one master and one slave. An IOPMP controlled by CVA6 filters master transactions. The cluster provides a DMA with one AXI4 port and 4 ports towards the L1SPM for high-bandwidth, low-latency transactions to/from the L1SPM. A dedicated event unit enables fine-grain parallel thread dispatching. Efficient communication between cluster and host domain is implemented through a dedicated hardware mailbox.

\vspace{-2mm}
\section{Software Stack and Programming Model}\label{Softwarestack}
\begin{figure}
    \centering
    \includegraphics[width=.5\textwidth]{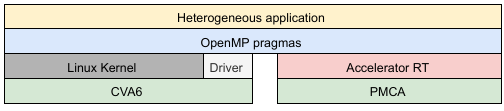}
    \caption{HULK-V SW stack (from HW to user-space)}
    \label{fig:openMP}
    \vspace{-4mm}
\end{figure}

HULK-V comes with a mature software stack for heterogeneous programming, as illustrated in Figure~\ref{fig:openMP}. Supporting Linux and an OpenMP-based framework for heterogeneous programming, HULK-V's software stack furnishes a well-known, popular, and mainstream set of user-level libraries, easing the programmability and enabling straightforward porting of legacy Linux-compatible applications.

On the PMCA side, HULK-V supports a lightweight bare-metal runtime that allows low programming overhead and is optimized for the small L1SPM. The PMCA runtime also allows hardware functionality validation and performance and power profiling. On the host side, CVA6 runs a full-fledged Buildroot-based Linux distribution (v5.16.9) equipped with a dedicated driver for the PMCA management but also supports a HULK-V bare-metal runtime.

CVA6's MMU supports SV39 virtual memory paging~\cite{privilegedriscv}, while the PMCA can only generate 32-bit addresses. A special main memory shared region, accessible through user-space \emph{hulk\_malloc()} function, enables data sharing in this mixed-address space. The function allocates contiguous memory buffers within accessible memory space, making the pointer sharing between the subsystems straightforward.

The APIs provided by the PMCA runtime and the Linux driver are already sufficient to run heterogeneous code on the platform. However, one must write two different codes for the host and cluster. To avoid this, HULK-V adapts the OpenMP 5 framework from HERO~\cite{herov2}, allowing users to use a high-level, directive-based, intuitive programming interface to efficiently offload the computationally intensive part of a program to the PMCA within one single heterogeneous source code. The HULK-V software stack comes with runtime libraries and compiler extensions of the Clang/LLVM 12 compiler for OpenMP 5 offload support from CVA6 ISA to RI5CY ISA.

\vspace{-3mm}
\section{Physical Implementation}\label{PHY}

\begin{figure}
    \centering
    \includegraphics[width=0.5\textwidth]{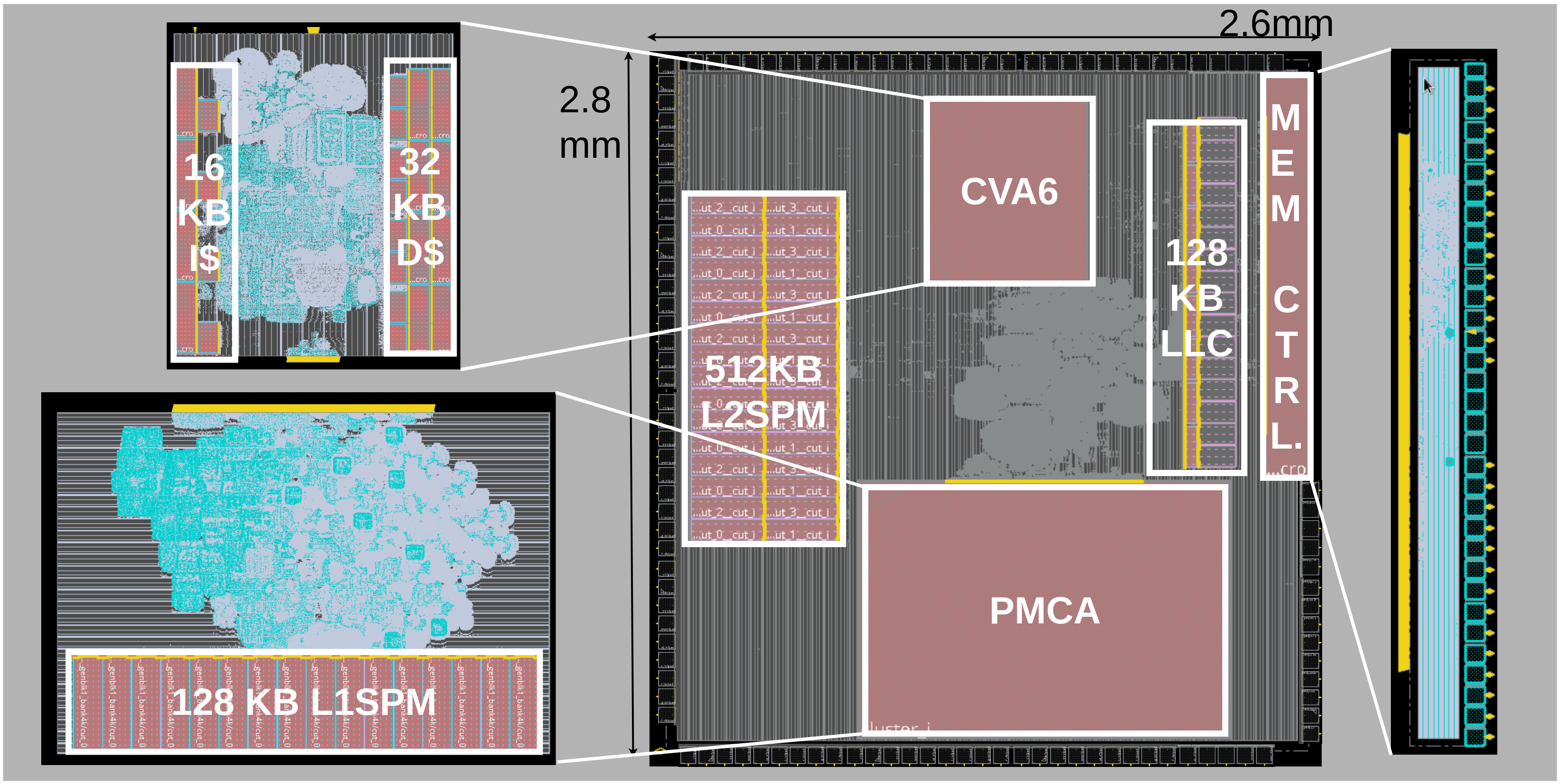}
    \caption{HULK-V floorplan.}
    \label{fig:layout}
    \vspace{-4mm}
\end{figure}

We implemented HULK-V in 22 nm FDX technology from Global Foundries, down to ready-for-silicon layout, with the aim of reliably estimating operating frequency, power, and area of the SoC. Physical synthesis has been performed with Synopsys Design Compiler, Place \& Route with Cadence Innovus, power analysis with Synopsys PrimeTime extracting value change dump (VCD) traces on the post layout, parasitics annotated netlist of the design with Siemens Questasim. Figure~\ref{fig:layout} shows the layout of the proposed SoC, featuring an area smaller than 9 mm$^2$.

CVA6 can reach up to 900 MHz in the worst corner (SSG corner at 0.72 V, -40/125 \degree C). The other components can reach between 400 and 450 MHz in the same corner. Table~\ref{tab:power_table} shows the power consumption in typical corner, at 0.8 V and 25\degree C. The HyperRAM controller consumes less than 2 mW at maximum frequency, around two orders of magnitude less than DDR controllers~\cite{powerhighend}. The overall power envelope of the SoC goes from 70 mW to 240 mW, depending on the active blocks in the system and their frequency.

\begin{table}[]
    \centering
    \caption{Power consumption at 25\degree C, 0.8V, TT}
    \begin{tabular}{|c|c|c|c|c|c|} \hline
                  & Area       & Leakage & Dynamic            & Max Freq & Max Power\\ 
                  & ($mm^2$)   & ($mW$)  & ($\frac{uW}{MHz}$) & (MHz)    &  (mW)    \\ \hline
         Top      & 7.28       & 4.23    & 214.7              & 450      & 100.53 \\\hline
         CVA6     & 0.49       & 4.79    & 47.5               & 900      & 47.54  \\ \hline
         PMCA  & 1.56       & 5.78    & 206                & 400      & 88.18  \\ \hline
         Mem Ctrl. & 0.27       & 0.14    & 2.3                & 450      & 1.16   \\ \hline
         Total    & 7.28       & 14.94   & 469.8              & -        & 237.41 \\ \hline
    \end{tabular}
    \label{tab:power_table}
    \vspace{-4mm}
\end{table}

\vspace{-3mm}
\section{Benchmarking}\label{BENCH}
We emulate HULK-V on the Xilinx VCU118 FPGA development board to measure performance. On FPGA, HULK-V can instantiate the HyperRAM controller or a proprietary Xilinx AXI4 DDR4 controller. While the DDR4s are already available on board, the HyperBUS is connected to one HyperRAM mounted on a PCB board plugged into the FMC connector. 

On FPGA, HULK-V runs at 50 MHz, the DDR4 controller interface at 165 MHz, and the DDR4 PHY at 1.2 GHz, while the HyperBUS runs at 25 MHz. The DDR4 models an ideal off-chip memory, faster by one order of magnitude than the SoC, while the HyperRAM works at half the frequency of the SoC, as it is for the ASIC SoC. Doing so allows us to sample the performance counters and obtain the same data on ASIC. On FPGA, we measure the operations per cycle. Combining Ops/Cycle and the measures obtained with Synopsys Primetime, we obtain the GOps and GOps/W of HULK-V with the components running at the frequencies listed in Table \ref{tab:power_table}.

\vspace{-4mm}
\subsection{Programmable Multi-Core Accelerator and CVA6}
To assess the performance and offloading overhead of the PMCA, we run a set of integer and floating-point DSP kernels on CVA6 and the cluster. All the benchmarks can be run on the cluster with reduced precision (FP16 for float and int8 for integer) to exploit the SIMD extensions unavailable on the CVA6 core and increase the operations per cycle.

\begin{figure}
    \centering
    \includegraphics[width=.5\textwidth]{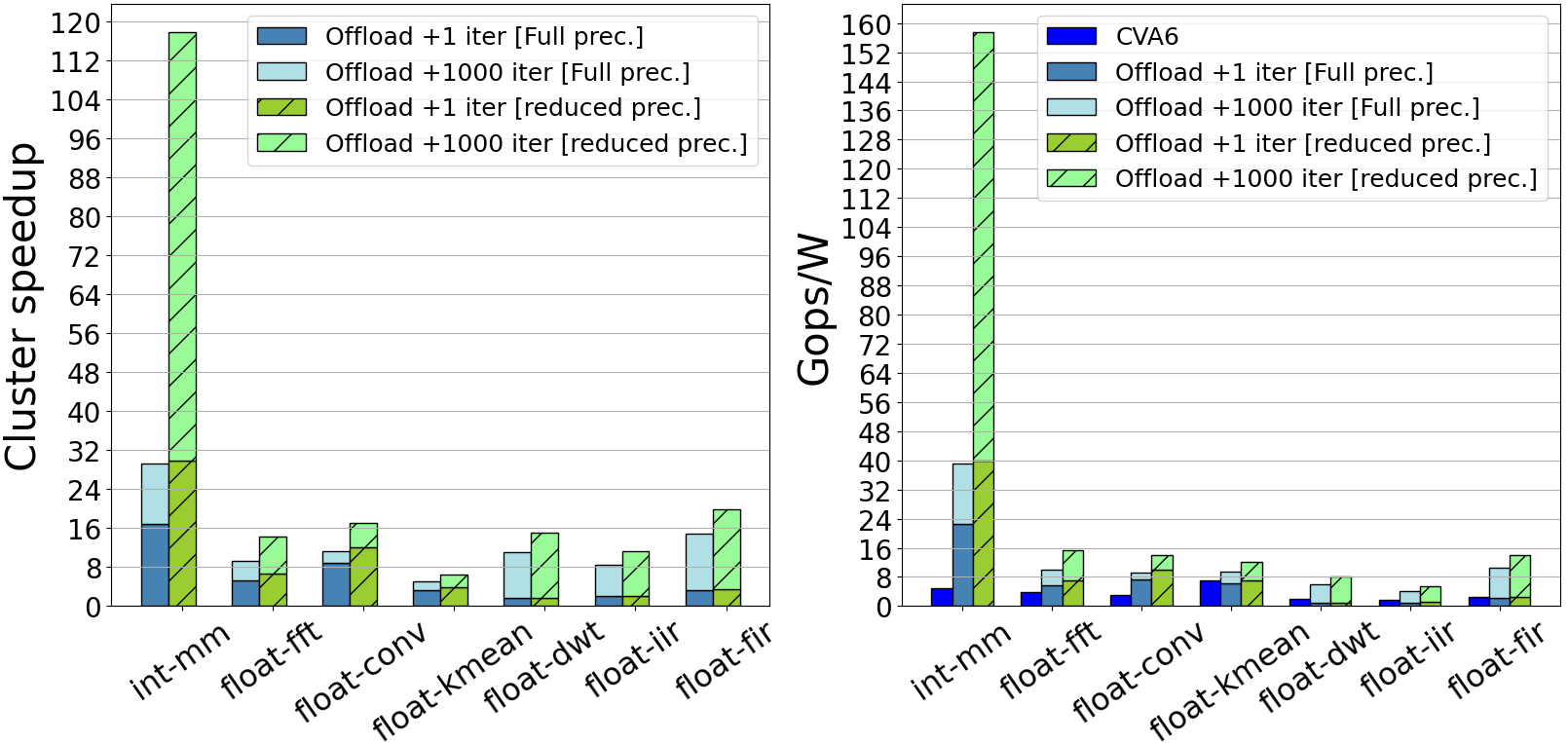}
    \caption{Speedup and Energy Eff. on PMCA vs CVA6.}
    \label{fig:cl_speedup}
    \vspace{-4mm}
\end{figure}

The left plot in Figure~\ref{fig:cl_speedup} shows the cluster's speedup in terms of the number of cycles. To estimate the two opposite boundaries in terms of code utilization, the figure shows the acceleration when executing the accelerated kernel once or 1000 times on the cluster. Due the fact that OpenMP offload triggers the load of the code \emph{lazily} (at first occurrence), when we run the inner kernel just once, if the kernel execution time is very short ($<$100k cycles), the cluster's offload overhead (i.e., loading the code into the L2SPM) dominates the total execution time and reduces the speedup. Luckily this is a very uncommon case, and even there, offloading to the cluster halves the execution time.

The right plot in Figure~\ref{fig:cl_speedup} shows the energy efficiency achieved by CVA6 and the cluster on the same benchmarks, with the IPs working at the maximum frequency. On a reduced-precision matrix multiplication, the cluster can reach up to 157 GOps/W, while CVA6 can only provide 4.9 GOps/W, 32$\times$ less. The matrix multiplication is representative of many ML applications, such as deep neural networks, and thanks to the high regularity and parallelizability, it is an easy target got the PMCA. On the other hand, the FP applications are less regular, and the minimum precision scales down to 16-bit and not 8-bit, being a more challenging target for the PMCA. Nevertheless, when executing the kernel many times, the PMCA can offer at least five times faster execution than CVA6 and higher energy efficiency on FP kernels.

\vspace{-4mm}
\subsection{Benchmarking of the Fully Digital Memory Hierarchy}

\begin{figure}
    \centering
    \includegraphics[width=.5\textwidth]{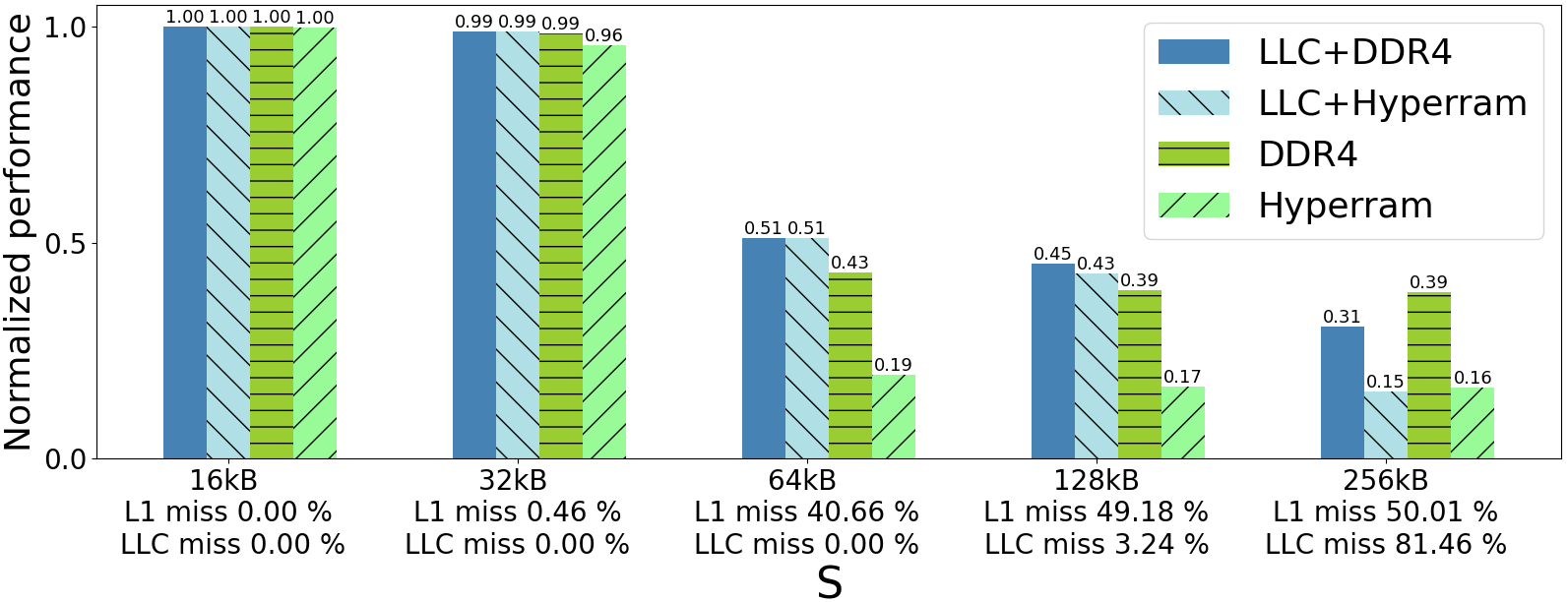}
    \caption{Sweep on Last Level Cache}
    \label{fig:llc_sweep}
    \vspace{-5mm}
\end{figure}

To benchmark the proposed lightweight, fully digital memory hierarchy with respect to a full-blown DDR one, we measure and compare CVA6's performance firstly on a synthetic benchmark and then on five IoT CPU-centric benchmarks, with four different memory configurations: 1) having the DDR4 and the LLC, 2) having the HyperRAMs and the LLC, 3) having just the DDR4 and 4) having just the HyperRAMs.

The synthetic benchmark, specifically designed to stress the cache hierarchy generating a controllable number of misses, consists of the following: we first read a whole 4 kB L1 cache way, the 0th, filling it. Then, we do many rounds of 4 kB reads with stride S. The second iteration warms up the caches. Within the remaining loops, reads can either be in the 0th way, causing either a miss or a hit, or in a different cache way and hit. The cache miss ratio increases with S.

\begin{figure}
    \centering
    \includegraphics[width=.5\textwidth]{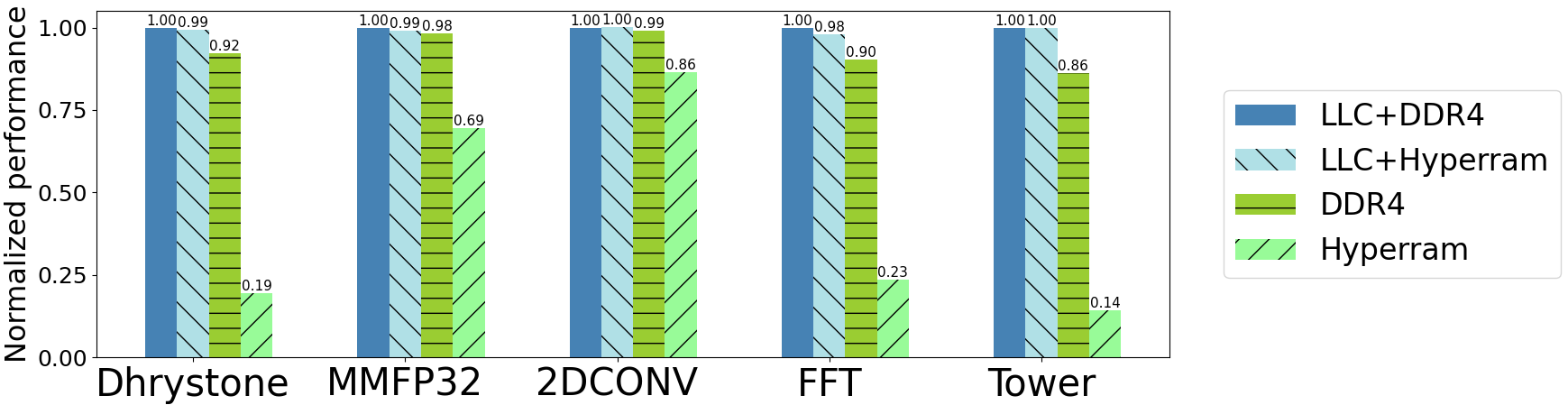}
    \caption{Last Level Cache effect}
    \label{fig:llc}
    \vspace{-4mm}
\end{figure}

Such a benchmark draws a lower performance bound: the resulting data pattern is highly unlikely to happen in real-world applications. Figure \ref{fig:llc_sweep} shows that CVA6's performance would not benefit from replacing the HyperRAMs with DDR4s when the L1 miss ratio is below 50\%, which is a reasonable assumption for the target embedded applications. Indeed, Figure~\ref{fig:llc} shows that the current cache hierarchy properly handles real-world IoT benchmarks. As expected, cases 1 and 2 have very similar performance, closer than 5\%, meaning that LPDDR/DDR memories would be oversized for our use cases.

\vspace{-4mm}
\subsection{Energy Efficiency Assessment of the Fully Digital Memory Hierarchy}
Once assessed the extremely limited performance degradation of the proposed memory hierarchy, we demonstrate its benefits in terms of energy efficiency, still comparing it with respect to a full-blown LPDDR4. We measure the computation-to-communication ratio of the benchmarks presented before plus two end-to-end DNNs (one for classification~\cite{dory} and one for autonomous navigation of drones~\cite{palossi201964}), exploiting DORY~\cite{dory} as memory-aware deployment flow, and Dhrystone. $CCR_{hyper}$ is defined as the ratio between the computing time and the time spent reading from the main memory, assuming full overlap of computation and communication phases, which is typical of explicitly memory-managed accelerators~\cite{dory}.

\begin{figure}
    \centering
    \includegraphics[width=.5\textwidth]{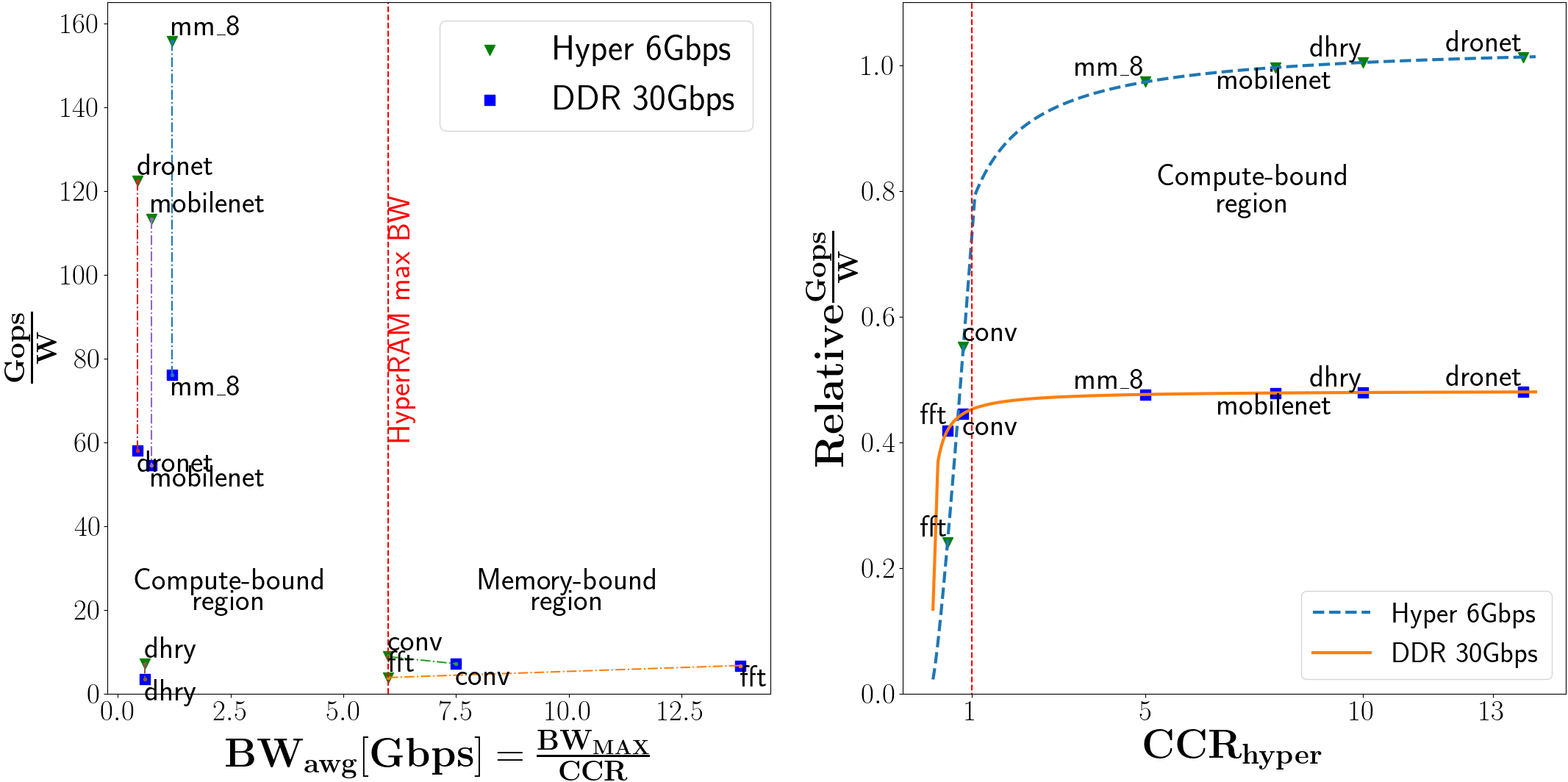}
    \caption{HULK-V Energy Efficiency.}
    \label{fig:power_s}
    \vspace{-4mm}
\end{figure}

The plot on the left in Figure~\ref{fig:power_s} shows the results. On the left of the line, there are compute-bound applications, that achieve maximum GOps with limited bandwidth. On the right of the line, there are memory-bound applications, benefitting from higher bandwidth in terms of GOps. However, that is not always the case also for energy efficiency, due to the much higher power consumption of the LPDDR4 and mixed-signal controller. To generalize, we also plot the relative energy efficiency against the $CCR_{hyper}$ in the right plot. Most of the IoT target applications, especially on the cluster, are compute-bound, thanks to the careful, deeply optimized data movements. For typical IoT applications with high data reuse, our memory hierarchy achieves double energy efficiency and the same performance as having an LPDDR4-based equivalent memory subsystem, while drastically reducing die area and cost.

\vspace{-4mm}
\section{Conclusion}\label{concoooo}
In this paper, we presented HULK-V: the first open-source low-cost heterogeneous RISC-V SoC running Linux within a 250 mW power envelope while providing up to 13.8 GOps and 157 GOps/W. Our memory subsystem enables Linux execution while keeping a small power factor; it provides comparable performance to equipping the board with LPDDR4 memories while achieving double GOps/W on ML parallel applications with high data reuse. Furthermore, our memory subsystem is open-source, fully-digital and cheap, occupying 0.27 $mm^2$ as compared to large (few $mm^2$ in the same technology node\cite{xeon}) proprietary mixed-signal PHY controllers. Finally, HULK-V provides a user-friendly OpenMP-based programming model for compiler-assisted heterogeneous code generation.



\end{document}